\input{aipcheck.tex}

\documentclass{aipproc}

\layoutstyle{8x11double}

\usepackage{graphicx}
\usepackage{epstopdf}

\def\lesssim{\mathrel{\rlap{\lower4pt\hbox{\hskip1pt$\sim$}}}<}
\def\gtrsim{\mathrel{\rlap{\lower4pt\hbox{\hskip1pt$\sim$}}}>}

\begin{document}

\title[Dark matter and the first stars] {The Effect of Dark matter on 
the first stars: a new phase of stellar evolution}

\author{Katherine Freese} {
address={Michigan Center for Theoretical Physics, University of Michigan, Ann Arbor, MI 48109},
,email={ktfreese@umich.edu]}}
\author{Paolo Gondolo} {
address={Physics Dept., University of Utah, Salt Lake City, UT 84112},
,email={paolo@physics.utah.edu}}
\author{Douglas Spolyar} {
address={Physics Dept., University of California, Santa Cruz, CA 95064},
,email={dspolyar@physics.ucsc.edu}}



\begin{abstract} 
Dark matter (DM)
in protostellar halos can dramatically alter the current
theoretical framework for the formation of the first stars.
Heat from supersymmetric DM annihilation can overwhelm any
cooling mechanism, consequently impeding
the star formation process and possibly leading to a new stellar phase.
The first stars to form in the universe may be
``dark stars'': giant 
($\gtrsim 1$ AU) hydrogen-helium stars powered by 
DM  annihilation instead of nuclear fusion. 
Possibilities for detecting dark stars are discussed.
\end{abstract}

\date{\today}

\maketitle

\section{Introduction}

At a lunch with David Gross, Director of KITP and winner of the 2004
Nobel Prize, one of us (K.F.) 
asked him what his goals were for the Large Hadron
Collider, the billion dollar accelerator at CERN in Geneva that 
will start taking data this spring.  His answer was, ``Supersymmetry,
Supersymmetry, Supersymmetry.''  Supersymmetry (SUSY), at this
point a beautiful theoretical construct, has the capability of addressing
many unanswered questions in particle theory as well as providing
the underpinnings of a more fundamental theory such as string theory.
If SUSY is right, then for every known particle in the universe, there
is an as yet undiscovered partner. The lightest of these, known
as the Lightest Supersymmetric Particle or LSP, would provide the dark
matter in the universe.  

The LSP is the favorite dark matter candidate of many physicists.
This is true not only because of the beautiful properties of SUSY,
but also because the LSP automatically has the right properties
to provide 24\% of the energy density of the universe.  In particular,
the neutralino, the SUSY partner of the W, Z, and Higgs bosons,
automatically has the required weak interaction cross section
and $\sim$ GeV - TeV mass to give the correct amount of dark matter in
the universe today.  The SUSY particles are in thermal equilibrium
in the early universe, and annihilate among themselves to produce
the relic density today.  It is this same annihilation process
that is the basis of the work we consider here.  The SUSY particles, also
known as WIMPs (Weakly Interacting Massive Particles),
annihilate with one another wherever their density is high enough.
Such high densities are achieved in the early universe, in galactic
haloes today \cite{ellis,gs}, 
in the Sun \cite{sos} and Earth \cite{freese,ksw}, 
and, as we have found,
also in the first stars \cite{sfg}. As our canonical values, we will 
use the standard value 
$\langle \sigma v \rangle = 3 \times 10^{-26} {\rm cm^3/sec}$ for the
annihilation cross section
and $m_\chi = 100$ GeV for the SUSY particle mass;
but will also consider a broader range of 
WIMP masses (1 GeV--10 TeV) and cross-sections.

We here describe the results of our work \cite{sfg} 
which considers the effect of SUSY dark matter annihilation
on the first stars.  These stars form at redshifts $z \sim 10-50$ 
in dark matter (DM) haloes
of $10^6 M_\odot$ (for reviews see e.g. 
\cite{Ripamonti:2005ri,Barkana:2000fd,Bromm:2003vv}).  
One star is thought to form inside one such
DM halo.  We must first ask, what is the dark matter density inside
a forming protostar?  To answer this question we
use an adiabatically contracted NFW profile
\cite{NFW}.  We start with an overdense region of $10^5-10^6 M_\odot$
with an NFW profile for both DM and gas, where the gas contribution is
15\% of that of the DM. Then we use adiabatic contraction ($M(r)r$ =
constant) \cite{Blumenthal:1985qy} and match
onto the baryon density profiles given by \cite {ABN,Gao06} to
obtain DM profiles.  Our resultant
DM profiles are shown in Fig.~1a for concentration parameter $c=10$ at
a redshift $z=19$ and halo mass $M=10^6 M_\odot$.  It is important
to point out that our results do not change much when these parameters
change; e.g. even for $c=1$, the dark matter density only changes
by a factor of 4.  After contraction, the DM density at
the outer edge of the baryonic core is roughly $ \rho_\chi \simeq 5
{\rm GeV/cm}^{-3} (n/{\rm cm}^{-3})^{0.81} $ and scales as $\rho_\chi
\propto r^{-1.9}$ outside the core.  
Our adiabatically contracted NFW profile matches
the DM profile obtained numerically in~\cite{ABN}, who
also found $\rho_\chi \propto r^{-1.9}$, for both their 
earliest and latest profiles.

\begin{figure}[t]
\centerline{\includegraphics[width=0.5\textwidth]{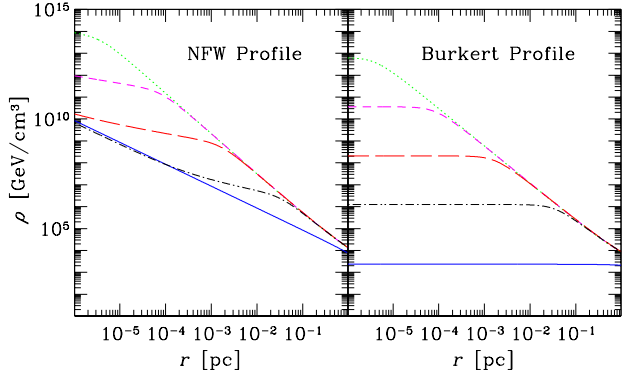}}
\caption{ Adiabatically contracted DM profiles for
(a) an initial NFW profile and (b) an initial Burkert profile \cite{burkert}, 
for $M_{\rm vir}=10^6 M_\odot$, $c_{\rm vir}=10$, and $z=19$.  The
blue (solid) lines show the initial profile.  Black (dot-dash) lines
correspond to a baryonic core density of $10^7{\rm cm}^{-3}$, red (long-dashed)
lines to $10^{10}{\rm cm}^{-3}$, magenta (dashed) lines to 
$10^{13} {\rm cm}^{-3}$ and green (dotted) lines to
$n\sim 10^{16}{\rm cm}^{-3}$.
\vspace{-\baselineskip}
}
\end{figure}

WIMP annihilation produces energy at a rate per unit volume $Q_{\rm
ann} = \langle \sigma v \rangle \rho_\chi^2/m_\chi \linebreak \simeq  1.2
\times  10^{-29} {\rm erg/cm^3/s} \,\,\, (\langle \sigma v \rangle / (3
\times 10^{-26} {\rm cm^3/s}))  \linebreak(n/{\rm cm^{-3}})^{1.6} (m_\chi/(100
{\rm GeV}))^{-1}$. In the early stages of Pop III star formation, when
the gas density is low, most of this energy is radiated away 
\cite{Ripamonti:2006gr,Chen:2003gz}. However,
as the gas collapses and its density increases, a substantial fraction
$f_Q$ of the annihilation energy is deposited into the gas, heating it
up at a rate $f_Q Q_{\rm ann}$ per unit volume.  We have estimated 
the fraction $f_Q$ of
DM annihilation energy that remains inside the gas core.  While neutrinos
escape from the cloud without depositing an appreciable amount of energy,
electrons and photons  can transmit energy to the core.

We find that, for LSP mass
$m_\chi = 100$GeV (1 GeV), a crucial transition takes place when
the gas density reaches $n> 10^{13} {\rm cm}^{-3}$ ($n>10^9
{\rm cm}^{-3}$).  Above this density, 
most of the annihilation energy remains inside the core and heats
it up to the point where further collapse of the core becomes difficult.
To compare with DM heating, we include all relevant cooling mechanisms.
The dominant mechanism is H$_2$ cooling; 
we use the rates in~\cite{Hollenbach}.
We use the opacities from \cite{Yoshida06}; e.g. at $n \sim 10^{13} {\rm
cm}^{-3}$, we take a $\sim 8\%$ cooling efficiency.
Setting the heating rate equal to the cooling rate gives the critical
temperature $T_c(n)$ at a given density $n$ below which heating dominates.  

In figure 2 we compare $T_c(n)$  to
typical evolution tracks 
in the temperature-density phase plane.  The blue (solid) and green 
(dotted) lines show
the temperature evolution of the protostellar gas in the
simulations (without DM) of \cite{Yoshida06} and \cite{Gao06} respectively.
The red (dashed and dot-dashed) lines show the critical temperature: 
below these lines, DM heating dominates over all cooling mechanisms.  

\begin{figure}[t]
\centerline{\includegraphics[width=0.5\textwidth]{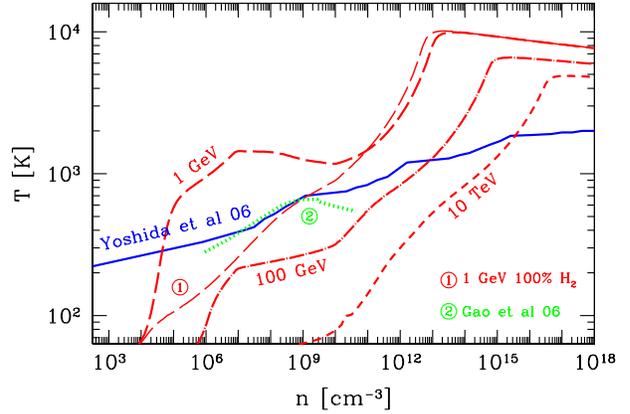}}
\caption{ Comparison of critical temperature  (red dashed lines) to
typical evolution tracks
in the temperature-density phase plane.  The blue (solid) and green 
(dotted) lines show the protostellar gas evolution from
simulations of
\cite{Yoshida06} and \cite{Gao06} respectively.
The red dashed lines mark $T_c(n)$ 
for: (i) $m_\chi = 1$ GeV with H$_2$ density from
simulations, (ii) $m_\chi = 1$ GeV assuming 100\% H$_2$, 
(iii) $m_\chi = 100$ GeV and (iv) $m_\chi  = 10$ TeV.
At the crossing point of the blue/green and red 
lines, DM heating dominates over cooling in the
core's evolution.
\vspace{-0.5\baselineskip}
}
\end{figure}

Figure 2 illustrates results for a range of WIMP masses from 
1 GeV--10 TeV for a canonical
$3 \times 10^{-26} {\rm cm^3}/{\rm sec}$ annihilation cross-section.
Since the heating rate scales as $\langle \sigma v \rangle / m_\chi$,
these same curves equivalently  
apply to a variety of cross-sections for a given WIMP mass.

The important result is that the blue/green (evolutionary) and red 
(critical temperature)
lines always cross, regardless of WIMP mass or H$_2$ fraction: this is a 
robust result.  As soon as the core density reaches this crossing point,
the DM heating dominates inside the core and changes its
evolution.  
Notice that at $m_\chi=$1 GeV, 
the crossing point for small ${H_2}$ fraction is at low densities, around 
$n \sim 10^5 {\rm cm}^{-3}$, in agreement with \cite{Ripamonti:2006gr}.
If the ${H_2}$ fraction is increased, cooling dominates 
for a longer time, as expected, but not forever.
Our results were obtained for two possible values for the H$_2$ fraction: 
the value given by the simulations without DM, and 
the case of 100\% molecular hydrogen. 

As soon as the DM annihilation products are contained inside the
protostellar core, the heating dominates over the cooling. Hence, for 100 GeV
(1 GeV) neutralino DM, 
once the gas density reaches a critical value of $\sim 10^{13} {\rm
cm}^{-3}$ ($10^9 {\rm cm}^{-3}$), the heating rate from DM
annihilation exceeds the rate of hydrogen cooling. The protostellar core is 
{\it
prevented from cooling and collapsing} further.  
The size of the core in the standard Pop III models
at this point is $\sim 17$ A.U. 
($\sim 960$ A.U.) and its mass is $\sim 0.6 M_\odot$ ($\sim 11  M_\odot$);
we plan to compute the stellar structures of the 
``dark stars'' to see what alternative size and mass result.
Our main conclusion is that the standard picture of Pop III star
formation is drastically modified by neutralino dark
matter annihilation inside the protostellar object.

We propose that a new type of object is created, a ``dark star''
supported by DM annihilation rather than fusion.  The question is
how long such a phase of stellar evolution lasts.  If such
an object were stable for a long time period, it would even be possible
for these dark stars to still exist today.  
Dark 
stars could last as long as 
the DM annihilation timescale,
$\tau_e = m_\chi/(\rho_\chi
\langle \sigma v \rangle) $ $\simeq$ $0.6~ {\rm Gyr}$  $(n/10^{13} {\rm
cm}^{-3})^{-0.8} $ $ (m_\chi/100 {\rm GeV}) $ $(\sigma v/3 \times
10^{-26}{\rm cm}^3 {\rm s}^{-1})^{-1}$. For our canonical case, we find $ \tau_e \sim 600$ Myr ($\sim 15$ Myr)  for $n = 10^{13} {\rm cm}^{-3}$ ($n = 10^{15}
{\rm cm}^{-3}$).
By comparison, the entire timescale for collapse (without
taking into account DM annihilation) is $\sim 1$ Myr at $z=50$ or
100 Myr at $z=15$; even for the more recent episodes, the
dynamical time at the high densities considered here is very short
($<10^3$ yr).  However, after this DM annihilates away, it is
possible that the DM hole in the small central core can fill in again,
depending on the DM orbits at this stage.  DM
further out can also continue to heat the core.   On the other
hand, as baryons continue to accrete onto the protostar, it is
possible that the annihilation shuts down sooner.  The lifetime
of the dark star phase is crucial to addressing the question
of the effects it has on the universe.

The effects of such a new phase of stellar evolution could be very
interesting.  The reionization of the IGM could be quite different,
as would be the production of the heavy elements required to form
all future generations of stars. DM heating may also alter the mass
of Pop III stars. Due to DM heating 
the initial mass function for Pop III stars could be modified. 
On the one hand the DM heating could prevent further accretion of baryons
\cite{mckee} so that the resulting stars are less massive.
Alternatively the initial protostellar object may be larger and
 dark stars might accrete enough material \cite{work} 
to form large black holes \cite{li,Pelupessy}
en route to building the $10^9 M_\odot$ black holes observed at $z \sim 6$.

What are the observational consequences of a ``dark star''?  
Dark stars
are giant objects, with core radii $\gtrsim 1 $ AU; perhaps they
could be found by lensing experiments. The DM
annihilation products may be seen in detectors on Earth today,
e.g. neutrinos travel great distances without interacting and might
be seen in e.g. AMANDA or ICECUBE. It is
interesting to imagine that DM could be discovered in this way. Or, if
it is previously discovered elsewhere, then its properties (mass and
cross-section) could be studied. The photons resulting from the
annihilation could contribute to the $\gamma$-ray
background and could be seen by GLAST or atmospheric 
Cherenkov telescopes such as HESS, VERITAS, and MAGIC. 
Alternatively, if the 
DM heating slows down but does not
entirely hinder the Pop III collapse, the difference of the expected
evolution from the more standard scenario could be seen in the next
generation of telescopes such as JWST.

\begin{theacknowledgments} 
This project would not have come into existence
without the help of Pierre Salati.  We are also particularly grateful
to Chris McKee for many useful conversations as well as his encouragement.
We also thank A. Aguirre, P. Madau, F. Palla, J. Primack,
R. Schneider, S. Stahler, G. Starkman, and N. Yoshida for discussions. 
K.F. acknowledges support
from the DOE and MCTP
via the Univ.\ of Michigan; from the Miller
Inst.\ at UC Berkeley; and thanks the 
Physics Dept.\ at UCSC.  D.S. and
K.F. thank the Galileo Galilei Inst. in Florence, Italy,
for support. P.G. acknowledges NSF grant PHY-0456825.
D.S. acknowledges NSF grant AST-0507117 and GAANN.
\end{theacknowledgments}


\begin{thebibliography}{9}


\bibitem{ellis}
J.R. Eliis, R.A. Flores, K. Freese, S. Ritz, D. Seckel, and J Silk,
Phys.\ Lett.\ B {\bf 214}, 403 (1988).

\bibitem{gs}
  P. Gondolo and J. Silk, Phys.\ Rev.\ Lett.\ {\bf 83}, 1719 (1999).

\bibitem{sos}
  M. Srednicki, K.A. Olive, and J. Silk, Nucl.\ Phys.\ B {\bf 279},
  804 (1987).

\bibitem{freese}
  K.~Freese,
  Phys.\ Lett.\  B {\bf 167}, 295 (1986).

\bibitem{ksw} 
  L.M. Krauss, M. Srednicki, and F. Wilczek, Phys.\ Rev.\ D {\bf 33},
  2079 (1986).

\bibitem{sfg}
  D. Spolyar, K.~Freese, and P.~Gondolo,
  astro-ph/0705.0521.

\bibitem{Ripamonti:2005ri}
  E.~Ripamonti and T.~Abel,
  astro-ph/0507130.

\bibitem{Barkana:2000fd}
  R.~Barkana and A.~Loeb,
  Phys.\ Rept.\  {\bf 349}, 125 (2001).

\bibitem{Bromm:2003vv}
  V.~Bromm and R.~B.~Larson,
  Ann.\ Rev.\ Astron.\ Astrophys.\  {\bf 42}, 79 (2004).

\bibitem{NFW}
  J.~F.~Navarro, C.~S.~Frenk and S.~D.~M.~White,
  Astrophys.\ J.\  {\bf 462}, 563 (1996).

\bibitem{Blumenthal:1985qy}
  G.~R.~Blumenthal et al., 
  Astrophys.\ J.\  {\bf 301}, 27 (1986).

\bibitem{ABN}
  T.~Abel, G.~L.~Bryan and M.~L.~Norman,
  Science {\bf 295}, 93 (2002).

\bibitem{Gao06}
 L.~Gao et al., 
  astro-ph/0610174.
  
\bibitem{burkert}
  A.~Burkert,
  IAU Symp.\  {\bf 171}, 175 (1996)
  [Astrophys.\ J.\  {\bf 447}, L25 (1995)].

\bibitem{Ripamonti:2006gr}
  E.~Ripamonti, M.~Mapelli and A.~Ferrara,
  Mon.\ Not.\ Roy.\ Astron.\ Soc.\  {\bf 375}, 1399 (2007).

\bibitem{Chen:2003gz}
  X.~L.~Chen and M.~Kamionkowski,
  Phys.\ Rev.\  D {\bf 70}, 043502 (2004).

\bibitem{Hollenbach}
  D.~Hollenbach and C.~F.~McKee,
   Astrophys.\ J.\  Suppl.\ {\bf 41}, 555 (1979).

\bibitem{Yoshida06}
  N.~Yoshida et al., 
  Astrophys.\ J.\  {\bf 652}, 6 (2006).

\bibitem{mckee} J.~C.~Tan and C.~F.~McKee,
  Astrophys.\ J.\  {\bf 603}, 383 (2004).
  
\bibitem{work}
Work in progress with N. Yoshida.

\bibitem{li}
  Y.~X.~Li {\it et al.},
  arXiv:astro-ph/0608190.

\bibitem{Pelupessy}
  F.~I.~Pelupessy, T.~Di Matteo and B.~Ciardi,
  arXiv:astro-ph/0703773.

\end{thebibliography}
\end{document}